\documentclass[twocolumn,showpacs,preprintnumbers,amsmath,amssymb]{revtex4}
\usepackage {longtable}
\usepackage{graphicx}
\usepackage{dcolumn}
\usepackage{bm}

\begin{document}

\title{Broadening and intensity redistribution in the Na($3p$) hyperfine excitation spectra due to optical pumping
in the weak excitation limit}

\author{I.~Sydoryk}

\author{N.~N.~Bezuglov}
\thanks{Fock Institute of Physics, St.Petersburg State
University, 198904 St. Petersburg, RUSSIA}

\author{I.~I.~Beterov}
\thanks{Institute of Semiconductor Physics SB RAS, 630090,
Novosibirsk, RUSSIA}

\author{K.~Miculis}

\author{E.~Saks}

\author{A.~Janovs}

\author{P.~Spels}

\author{A.~Ekers}

\affiliation{Laser Centre, University of Latvia, LV-1002 Riga,
LATVIA}

\date{February 6, 2008}

\begin{abstract}

Detailed analysis of spectral line broadening and variations in relative intensities of hyperfine spectral components
due to optical pumping is presented. Hyperfine levels of sodium $3p_{1/2}$ and $3p_{3/2}$ levels are selectively
excited in a supersonic beam at various laser intensities under the conditions when optical pumping time is
shorter than transit time of atoms through the laser beam. The excitation spectra exhibit significant line
broadening at laser intensities well below the saturation intensity, and redistribution of intensities of
hyperfine spectral components is observed, which in some cases is contradicting with intuitive expectations.
Theoretical analysis of the dynamics of optical pumping shows that spectral line broadening depends sensitively
on branching coefficient of the laser-driven transition. Analytical expressions for branching ratio dependent
critical Rabi frequency and critical laser intensity are derived, which give the threshold for onset of
noticeable line broadening by optical pumping. The critical laser intensity has its smallest value for
transitions with branching coefficient equal to 0.5, and it
can be much smaller than the saturation intensity. Transitions with larger and smaller branching
coefficients are relatively less affected. The theoretical excitation spectra were
calculated numerically by solving density matrix equations of motion using the split propagation technique, and
they well reproduce the observed effects of line broadening and peak intensity variations. The calculations also
show that presence of dark (i.e., not laser- coupled) Zeeeman sublevels in the lower state results in effective
branching coefficients which vary with laser intensity and differ from those implied by the sum rules, and this
can lead to peculiar changes in peak ratios of hyperfine components of the spectra.
\end{abstract}

 \maketitle



\section{Introduction}

Optical pumping is a well known phenomenon, which is usually
associated with redistribution of population within hyperfine (HF)
components or Zeeman sublevels of the ground state due to coupling
by resonant light fields \cite{Happer}. Optical pumping is being
exploited in various applications, like cooling below the Doppler
limit \cite{Tannoudji2, Metcalf}, vibrational excitation of
molecules in the electronic ground state \cite{Bergmann},
orientation and alignment of atomic and molecular ground states
\cite{Auzinsh}, etc. When optical pumping is involved in the
control of quantum states, it is usually associated with large
laser intensities exceeding the saturation limit \cite{Metcalf}.
Therefore, the populations of quantum states depend nonlinearly on
laser intensities and the excitation spectra are affected by power
broadening \cite{Demtroder}.

In present study we are concerned with lineshape effects due to
optical pumping in the weak excitation limit. Specifically, we
measure laser excitation spectra of the 3$p_{1/2}$ and 3$p_{3/2}$
states of Na in a supersonic beam. Coupling of the $F''=1$ and
$F''=2$ levels of the ground state with different HF components of
the upper states allows us to study two-level systems with
different branching coefficients. The smaller the branching
coefficient, the more population irreversibly leaves the two-level
system, and vice versa. At very low laser intensities the
excitation spectra do not reveal any abnormalities. When laser
intensity is increased but still below the saturation intensity,
essential modification of the excitation spectra is observed: most
of the hyperfine spectral components exhibit additional broadening
while their intensity ratios cease to obey the line strengths
rules. Note, that usually line broadening is considered to be a
strong-field effect due to power broadening at laser intensities
above the saturation intensity \cite{Metcalf, Demtroder}.


\begin{figure}
\label{fig1}
\includegraphics[width=8cm]{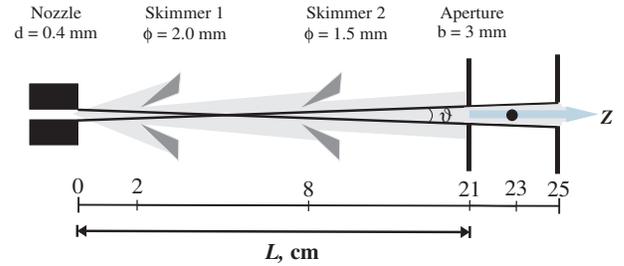}
\caption{Collimation of the sodium beam by skimmers and apertures.
Laser beam crosses the atomic beam at right angles 25~cm
downstream from the nozzle.}
\end{figure}


Since the experiments were performed at low number densities of
sodium atoms ($n_{3s}\sim $10$^{10}$cm$^{-3}$), lineshape
modifications by radiation trapping can be disregarded
\cite{Ivanov, BezAig}. We attribute the observed lineshape effects
to optical pumping, which leads to depletion broadening of
spectral lines \cite{Dep1, Dep2}. If transit time $\tau_{tr}$ of
atoms through the laser beam is much larger than lifetime
$\tau_{nat}$ of the 3$p$ state, populations of levels and the
associated fluorescence signals can become nonlinear on laser
intensity $I_{las}$ long before the saturation limit is reached
(i.e., at $I_{las}\ll I_{sat}$). Due to interaction with laser
field the ground state $g$ has a finite width
\cite{Tannoudji2,Tannoudji1}

\begin{equation}
\label{eq1}
\Gamma '' = \Omega ^{2}\frac{{\Gamma _{nat}} }{{\Gamma _{nat}^{2} + 4\delta
^{2}}},
\end{equation}

\noindent where $\Gamma _{nat} = 2\pi \cdot \Delta \nu _{nat} =
1/\tau _{nat}$ is the natural width (in units of angular
frequency [sec$^{-1}$]) of the excited state $e$, $\delta$ is
the laser detuning from the line center, and $\Omega $ is the
Rabi frequency of the transition. The width $\Gamma ''$ is
equal to the rate of photons spontaneously emitted from state
$e$. If the level system is partially open, only the fraction
$\Pi$ of the spontaneous transitions will return the population
to the initial state $g$; the fraction $1 - \Pi$ associated
with decay to levels other than $g$ will be lost from the
($g,e$)-system during each excitation-emission cycle. The rate
of such pumping is obviously $\Gamma_{pump}=\left( {1 - \Pi}
\right)\Gamma ''$. Hence, the pumping time can be written as

\begin{equation}
\label{eq2}
\tau _{pump} \left( {\delta}  \right) = \frac{{1}}{{\Gamma _{pump}} } =
\frac{{\Gamma _{nat}^{2} + 4\delta ^{2}}}{{\Gamma _{nat} \Omega^{2}
\left( {1 - \Pi}  \right)}}.
\end{equation}

\noindent If the transit time $\tau_{tr}$ is long, such that
$\tau_{tr}>\tau_{pump}^{(0)}\equiv \tau_{pump}(\delta=0)$, the population
of the ($g,e$)-system will be fully depleted during interaction with the laser field.
In terms of Rabi frequencies the condition for population depletion can be rewritten
as $\Omega>\Omega_{cr}\cong
\Omega_{sat}\sqrt
{{{2\tau_{nat}}\mathord{\left/{\vphantom{{2\tau _{nat}}{\left(
{\tau_{tr}\left({1 -
\Pi}\right)}\right)}}}\right.\kern-\nulldelimiterspace}{\left(
{\tau_{tr}\left({1 - \Pi}\right)}\right)}}}$, where saturation Rabi
frequency is $\Omega _{sat} = \Gamma _{nat} /\sqrt {2}$
\cite{Metcalf}. Note, that the parameter
$1/\sqrt {\tau _{nat} \tau _{tr}}$, which was considered in
\cite{Auzinsh, Ruth} as the parameter associated with saturation due to optical pumping
in the case of open level systems (i.e., no population return form state $e$ to state $g$),
is identical to our critical Rabi frequency $\Omega _{cr}$ in the limiting case of $\Pi$=0.

If the weak excitation limit is combined with long interaction
times of atoms with the laser field, such that $\tau _{tr} >> \tau
_{nat}$, the value of critical Rabi frequency is small
($\Omega_{cr}<\Omega_{sat}$) and broadening and saturation of
spectral lines can be observed at laser intensities well below the
saturation limit, long before power broadening starts affecting
the lineshapes.




\section{Experiment}


\begin{figure}
\label{Fig2}
\includegraphics[width=7cm]{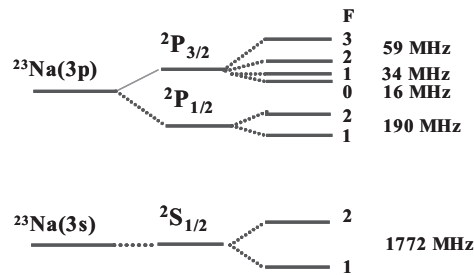}
\caption{Hyperfine energy levels of the 3$s$ and 3$p$ states of Na.}
\end{figure}



\begin{figure}
\label{Fig3}
\includegraphics[width=7cm]{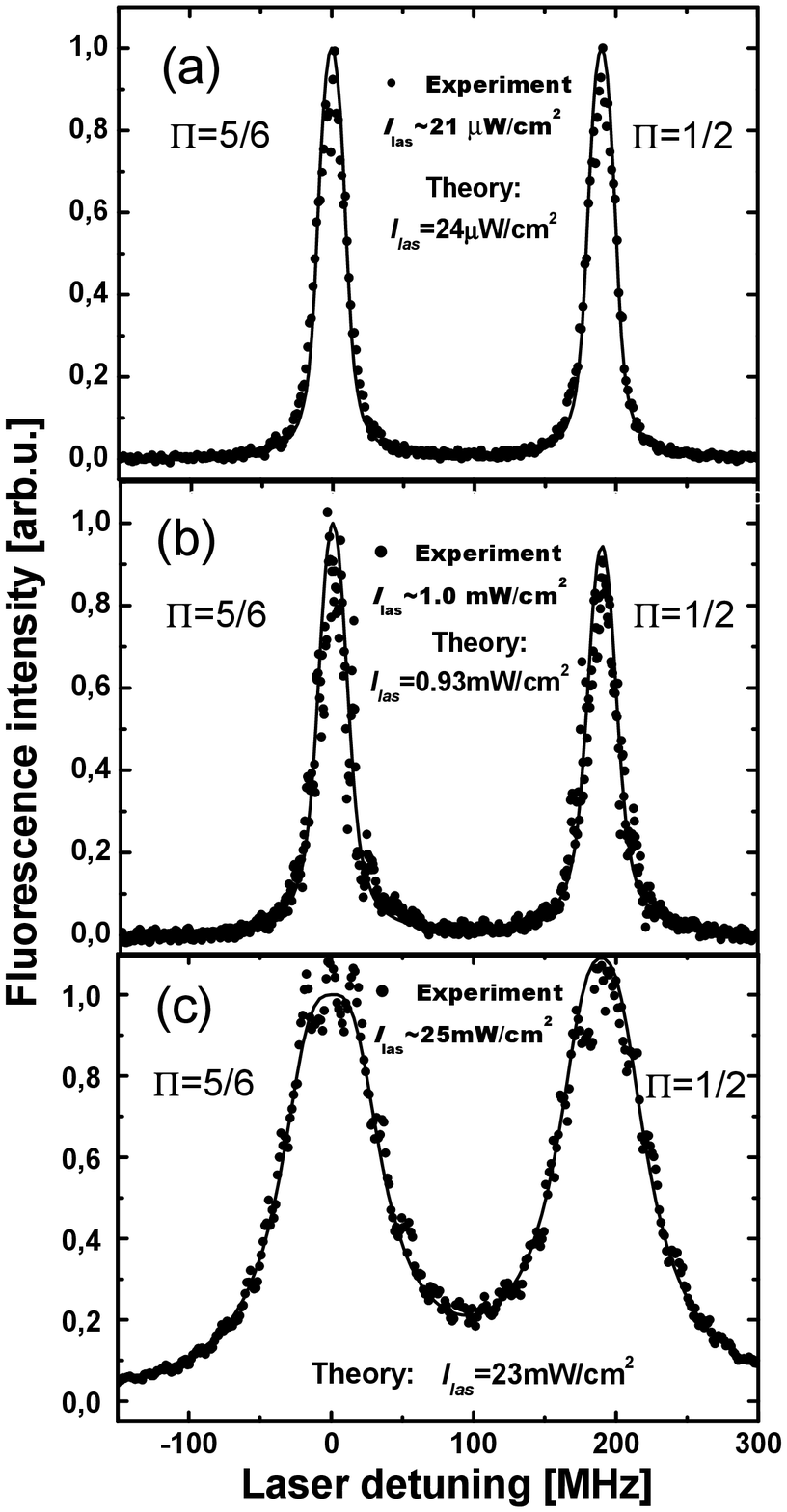}
\caption{Excitation spectra of the $3s_{1/2},F'' = 2 \to
3p_{1/2},F' = 1,2$ transitions in Na. Residual Doppler width due
to finite collimation angle is $\Delta \nu _{D}$=11.2 MHz (at
$b=2$~mm). The expected peak ratio is 1:1. Saturation intensities
of the lhs and rhs components are 7.5 and 12.5~mW/cm$^{2}$,
respectively.}
\end{figure}

The experiment was performed in a supersonic beam of Na atoms (see
Fig.~1). Two skimmers and an entrance aperture of the excitation
zone collimate the beam with flow velocity $v_f$ to a small
divergence angle $\vartheta$, thus reducing the Doppler width for
excitation perpendicular to the beam axis to $\Delta\nu_D^{}\equiv
v_f \vartheta/(2 \lambda)$, where $\lambda$ is the laser
wavelength. The divergence angle $\vartheta$ was varied between
$0.67^{\circ}$ and $0.92^{\circ}$ by using the entrance aperture
$b$ of either 2~mm or 3~mm dia. The laser beam crosses the atomic
beam at right angles, and it is linearly polarized parallel to the
molecular beam axis $z$, which is also the quantization axis. Only
Zeeman sublevels with identical quantum numbers $m_F$ are coupled
by the laser field due to the selection rule $\Delta m_F=0$, while
for transitions between levels with the same $F$ the transition
$m_{F''}=0\leftrightarrow m_{F'}=0$ is forbidden. The number
density of atoms in the beam was chosen sufficiently low
($\leq10^{10}$ cm$^{-3}$), thus ensuring that the beam is not
optically thick and effects of radiation trapping and photon
reabsorption \cite{BezAig} can be safely neglected. An important
consequence of optical transparency of the beam is that absorption
$P\left(\Delta \nu_L^{}\right)$ (the total number of photons
absorbed per second) and excitation $J\left( {\Delta \nu_L^{}}
\right)$ (the integrated over frequencies flux of emitted photons
in the direction of observation) profiles as function of the laser
detuning $\Delta \nu_L$ do not vary in the interaction volume
defined by the crossing atomic and laser beams. Both profiles are
proportional to the integral (over the interaction volume and HF
sublevels) population of the excited state.

The $3s\rightarrow 3p$ transition was excited using a single mode
cw radiation source (Coherent CR-699-21 dye laser) with linewidth
of 1MHz. The fluorescence emitted by Na atoms was collected into
two fiber bundles at the angles of $90^\circ$ and $45^\circ$ with
respect to the directions of the axis of the molecular beam, laser
beam, and laser polarization. The fluorescence light was guided
via the fiber bundles to two photomultipliers, and the signals
proportional to $J\left({\Delta \nu_L^{}}\right)$ were registered
using photon counters. The resulting excitation spectra of the
$3p$ state were recorded as a function of laser detuning $\Delta
\nu_L$. The arrangement with two different simultaneous detection
geometries allowed us to verify that radiation trapping, which is
strongly anisotropic with respect to the direction of the
observation, does not affect the measured spectra. It also allowed
us to rule out the influence of polarization effects on variations
in lineshapes and relative line intensities.

The mean flow velocity $v_{f}$ of atoms in the beam was measured
to be 1160~m/s. For excitation perpendicular to the atomic beam
axis the apertures of the excitation zone of $b=2$ and 3 mm
correspond to the residual Doppler width $\Delta \nu _{D}= 11.2$
and 15.9 MHz (FWHM), respectively. These should be compared to the
excitation perpendicular to the natural width of $\Delta \nu
_{nat}$ of 9.8~MHz ($\tau _{nat} $=16.23~ns for the $3p_{3/2}$
\cite{Jones}).

Figure~2 shows the hyperfine energy levels of the $3s$ and $3p$
states. The excitation spectra were obtained by scanning the laser
frequency across the $3s_{1/2}\rightarrow 3p_{1/2}$ and
$3s_{1/2}\rightarrow 3p_{3/2}$ transitions. The HF splittings are
larger than both the Doppler width and the natural width for all
but one pair of components ($3p_{3/2}$ $F'=0$ and $F'=1$). The
measurements for the D$_{1}$-line ($\lambda $=589.593~nm) were
performed with the aperture $b$=2~mm. In the case of the
D$_{2}$-line ($\lambda$=588.996~nm), $b$=3~mm was used. Radius of
the laser beam was $r_{las} $=1.5~mm, which corresponds to the
transit time $\tau _{tr}=2\times r_{las}/v_{f}$=2.65~$\mu$s at
$v_{f}$=1160~m/s. Thus, the transit time is by more than two
orders of magnitude larger than the natural lifetime of the $3p$
state.

\begin{figure}
\label{Fig4}
\includegraphics[width=7cm]{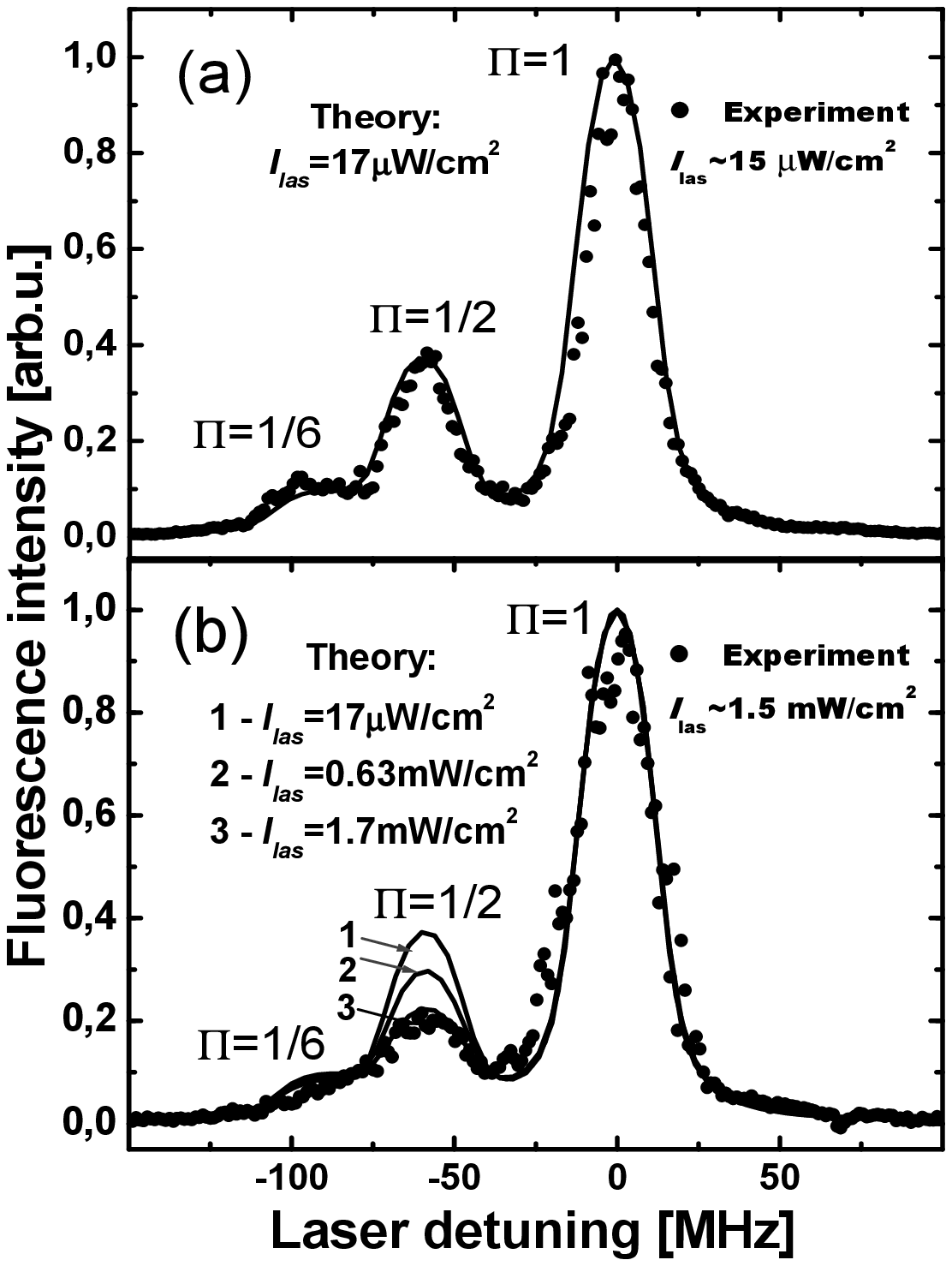}
\caption{Excitation spectra of the $3s_{1/2},F'' = 2 \to
3p_{3/2},F' = 1,2,3$ transitions in Na. Residual Doppler width due
to finite collimation angle is $\Delta \nu _{D}$=15.9 MHz (at
$b=3$~mm). The expected peak ratio is 1:5:14. Saturation
intensities of the three components are 37.4, 12.5, and
6.2~mW/cm$^{2}$, respectively.}
\end{figure}

\begin{figure}
\label{Fig5}
\includegraphics[width=7cm]{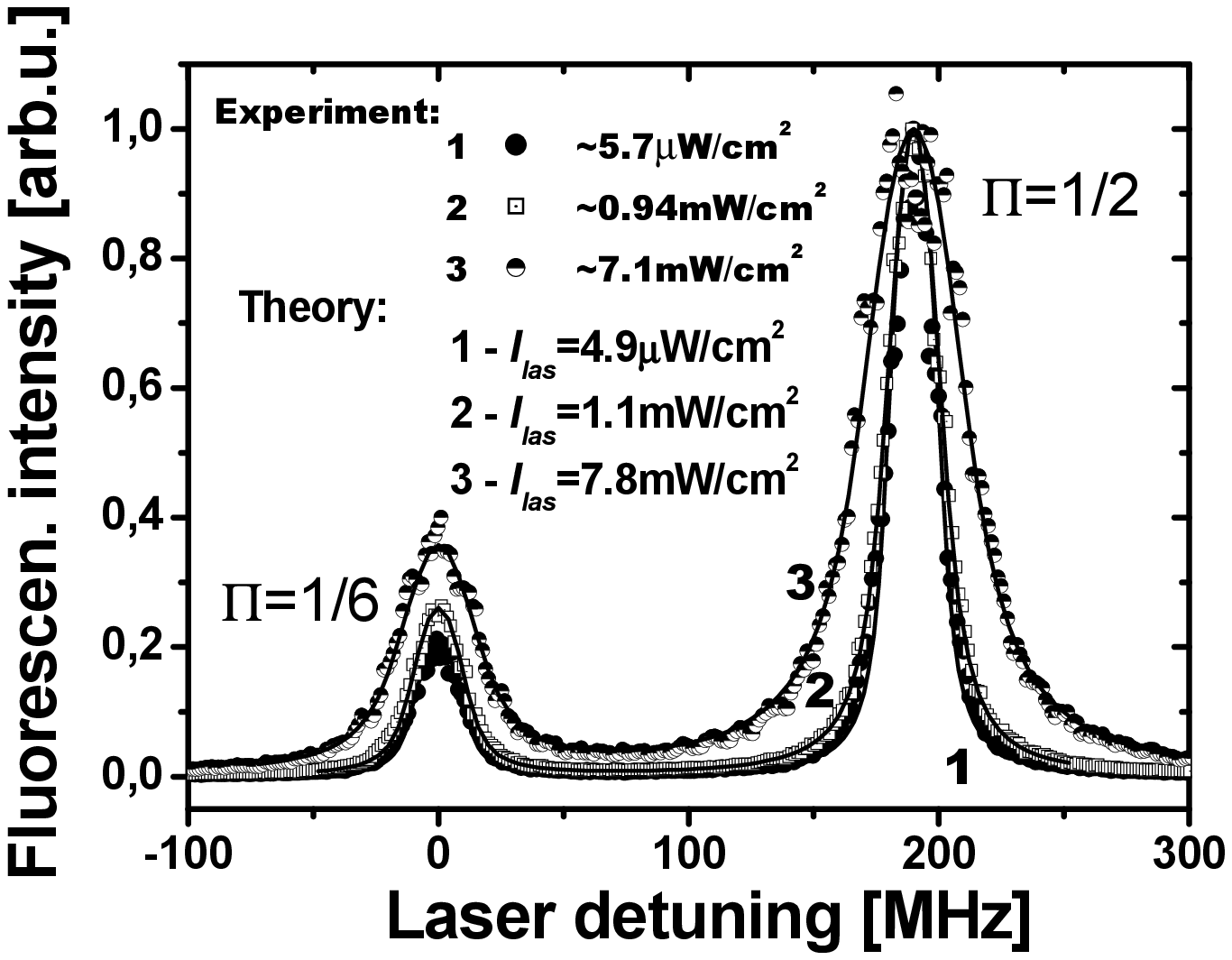}
\caption{Excitation spectra of the $3s_{1/2},F'' = 1 \to
3p_{1/2},F' = 1,2$ transitions in Na. Residual Doppler width due
to finite collimation angle is $\Delta \nu _{D}$=11.2 MHz (at
$b=2$~mm). The expected peak ratio is 1:5. Saturation intensities
of the lhs and rhs components are 37.4 and  12.5~mW/cm$^{2}$,
respectively.}
\end{figure}


\section{Spectra}

Figure~3 shows the measured excitation spectra of the $3p_{1/2}$
state from the $F''$=2 sublevel of the ground state at various
laser intensities. The spectra exhibit two peaks corresponding to
the excitation of the $3s_{1/2},F'' = 2 \to 3p_{1/2}, F' = 1,2$ HF
transitions. The spectra of Fig.~3 were measured at the divergence
angle of the atomic beam of $\vartheta = 0.67^{\circ}$, which
corresponds to residual Doppler width of $\Delta \nu_D = 11.2$~MHz
(see Sec.~VI for details on Dopller lineshape). The spectrum of
Fig.~3(a) was measured at a very low laser intensity of
$21\mu$W/cm$^2$. Both HF components appear equally strong, which
is obviously due to equal line strengths of both HF transitions
\cite{Sobelman}. The lineshapes are determined by a combined
effect of natural ($\Delta \nu_{nat}=9.8$~MHz) and Doppler
broadening. In Fig.~3(b) the laser intensity has been increased by
a factor of about 70 compared to Fig.~3(a) to the value of
1~mW/cm$^2$, which is still much smaller than the saturation
intensity of both HF components (7.5 and 12.5~mW/cm$^2$). One can
observe that the $F'' = 2 \to F' = 2$ (rhs) component has become
somewhat smaller than the $F'' = 2 \to F' = 1$ (lhs) component.
Peculiarly, when the laser intensity is further increased to
25~mW/cm$^2$ [Fig.~3(c)], the rhs component becomes somewhat
larger than the lhs component, while the the widths of the peaks
($\Delta \nu = 75$~MHz) are substantially larger than the width
$\Delta \nu_{sat} = 16.5$~MHz expected from saturation broadening
at this laser intensity.

\begin{figure}
\label{Fig6}
\includegraphics[width=7cm]{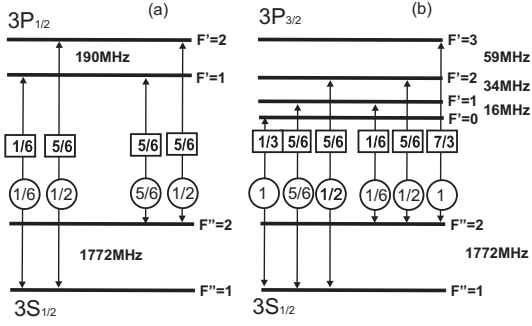}
\caption{Line strengths $\tilde {S}_{i}^{(j)}$ (square frames) and
branching ratios $\Pi_{i}$ (circular frames) for (a) $3s_{1/2} \to
3p_{1/2}$ and (b) $3s_{1/2} \to 3p_{3/2}$ hyperfine transitions.}
\end{figure}


The $3s_{1/2}, F''=2 \to 3p_{3/2}, F' = 1,2,3$ excitation spectra
are shown in Fig.~4. The relative peak intensities match the
theoretical line strengths of individual HF transitions when laser
intensity is very small [15$\mu$W/cm$^2$, Fig.~4(a)]. When laser
intensity is increased to 1.5$\mu$W/cm$^2$, which is still below
the saturation intensity, the relative intensities of the
components corresponding to the excitation of the $F'=1$ and
$F'=2$ HF levels are smaller than expected from the theoretical
line strengths [Fig.~4(b)]. When laser intensity is close to
saturation intensity, the $F'=1$ and $F'=2$ peaks are so weak
compared to the $F'=3$ peak that it is ambiguous to attempt
analysis of their linewidth.

The $3s_{1/2}, F''=1  \to 3p_{1/2}, F'=1,2$ excitation spectra are
shown in Fig.~5. Like in the case of Fig.~3, also here a
significant broadening is observed at laser intensities below the
saturation limit. In contrast to Fig.~3(b), however, the lhs peak
corresponding to the excitation of the F'=1 component of the upper
state grows monotonically as compared to the rhs peak.


\section{Theoretical line strengths and saturation intensity}

At very small laser intensities the strengths of individual peaks
in the excitation spectra shown in Figs.~3-5 correspond to the
respective theoretical line strengths $S_{i}^{\left( {j} \right)}$
of individual HF transitions $i = \{ F'' \to F'\} $ within the
D$_{1}$ ($j=1/2$) or D$_{2}$ ($j=3/2$) lines \cite{Sobelman}. The
values of $S_{i}^{\left( {j} \right)}$ are directly related to the
reduced matrix elements of the transitions and the partial natural
width of the respective transition:

\begin{equation}
\label{eq3} S_{i}^{\left( {j} \right)} = |\left(
{1/2,F''||D||j,F'} \right)|^{2}; \quad \Gamma _{nat}^{\left(
{i} \right)} = \frac{{4\omega _{i}^{3}} }{{3\hbar
c^{3}}}\frac{{1}}{{2F' + 1}}S_{i}^{\left( {j} \right)}.
\end{equation}

\noindent Intensity of the component $i$ is proportional to its
line strength because the product $\Gamma _{nat}^{\left( {i}
\right)} \left( {2F' + 1} \right)$ regulates the photon flux of
this component under the conditions of thermodynamic equilibrium
\cite{Sobelman}. Figure~6 shows the theoretical line strengths $
\tilde {S}_{i}^{\left( {j} \right)}$ (square frames) in units of
the reduced matrix element $|\left( {3s||D||3p}\right)|^{2}$ of
unresolved $3s \to 3p$ transition, i.e., $S_{i}^{\left( {j}
\right)} = \tilde {S}_{i}^{\left( {j} \right)} |\left( {3s||D||3p}
\right)|^{2}$. The values of $\tilde{S}_i^{\left( {j} \right)}$
are normalized such that $\Sigma _{j,i} \tilde{S}_i^{\left( {j}
\right)}=8$ \cite{Sobelman}. Thus, the theoretical values of peak
ratios can be directly taken from Fig.~6, and they agree with the
experimental observations at very small laser intensities [see
Figs.~3(a), 4(a), and 5].

Saturation intensity of each hyperfine transition depends on the
natural width of the transition $\Delta \nu_{nat}$ and the
branching ratio $\Pi _{i}$ \cite{Metcalf} (see also in Sec.~V):

\begin{equation}
\label{eq4} I_{sat}^{\left( {i} \right)} = \frac{{4\pi ^{3}\hbar
c}}{{3\lambda _{i}^{3} }}\frac{{\Delta \nu _{nat}} }{{\Pi _{i}} }.
\end{equation}

\noindent The values of the hyperfine branching coefficients $\Pi
_{i}$ (circular frames in Fig.~6) are easily obtained from the
reduced line strengths $\tilde {S}_{i}^{\left( {j} \right)}$.
Note, that all HF transitions have the same natural width of
$\Delta \nu _{nat} $=9.8~MHz.

The saturation intensity given by Eq. (\ref{eq4}) gives the
limiting laser intensity after which stimulated transitions start
transforming the excitation spectra \cite{Metcalf}. The lowest
laser intensities used in measurements of the spectra shown in
Figs.~3-5 are much smaller than saturation intensity
$I_{sat}^{\left( {i} \right)}$ of any of the HF transitions.
Therefore one naturally expects the peak ratios to be in
accordance with the line strengths of Fig.~6 and line width to
correspond to residual Doppler width determined by beam
divergence. This agrees with the observations made for the
smallest laser intensities.

When laser intensity is increased by a factor of about 100, it is
still well below the saturation intensity $I_{sat}^{\left( {i}
\right)}$. Nevertheless, a curious transformation of the spectra
is observed: widths of the peaks increase, and peak ratios of the
HF components change. Interestingly, the strongest $
F''=2\rightarrow F'=3$ component of the D$_{2}$-line (Fig.~4) is
not affected by broadening at all although its saturation
intensity $I_{sat}^{\left( {i} \right)} $=6.2 mW/cm$^{2}$ is the
smallest. Intuitively, one would expect the transition
$F''=2\rightarrow F'=3$ to be the first one that is affected by
broadening when laser intensity is increased.

Not only the widths are affected. Relative intensities of the HF
peaks change as laser intensity is increased. At the first glance
it seems that relative intensities of components with smaller
branching ratios $\Pi_i$ should decrease when optical pumping
becomes non-negligible, as less population returns to the lower
laser coupled level than it does for levels with larger $\Pi_i$.
This is clearly the case in Figs.~3(b) and 4, but not in the case
of Fig.~5. Moreover, Fig.~3 shows another unexpected feature:
after the intensity of the peak with smaller $ \Pi_i$ has
initially decreased with respect to the peak with larger $\Pi_i$
[cf. Figs.~3(a) and 3(b)], a further increase of laser intensity
leads to increase of the peak with smaller $\Pi_i$ [cf. Figs.~3(b)
and 3(c)]. Explanation of these observations requires a detailed
analysis of the dynamics of optical pumping.




\section{Dynamics of optical pumping and its effect on the fluorescence signals}

The measured fluorescence signals are affected by various factors, like the detection
efficiency and geometry. The spectral
components are excited and detected at very close wavelengths under identical conditions,
therefore it can safely assumed that the detection efficiency is equal for all of them.
Since we are interested in relative intensities and widths of the components, it is sufficient
to consider the fluorescence signals that are proportional to the total number of photons
emitted by atoms at all times in all directions.

We consider the following model problem. Two-level atoms with
the ground state $ g$ and the excited state $e$ propagate along
the $z$-axis with the flow velocity of the beam $v_{f}$.
The atoms cross the laser beam with radius $r_{las}$,
frequency $\omega $ and Gaussian intensity distribution

\begin{equation}
\label{eq5} I \left( {z} \right) = I_{las}
\mathrm{\exp}\left( { - z^{2}/r_{las}^{2}}  \right) \; ; \quad
\tau _{tr} = 2r_{las} /v_{f} \; .
\end{equation}

\noindent It corresponds to Gaussian switching of Rabi frequency $
\Omega$ of the $g-e$ transition:

\begin{equation}
\label{eq6} \Omega\left( {t} \right) = \Omega _{0}
\mathrm{\exp}\left( { - 2t^{2}/\tau _{tr}^{2}}  \right) \; ;
\quad t=z/\mathrm{v_f}.
\end{equation}

\noindent The value $\Omega _{0} = E_{0} \langle g|d_{z} |e\rangle
$ is the Rabi frequency of the $ g-e$ coupling in the center of
the laser beam, which is linearly polarized parallel to $z$-axis.


\subsection{Evaluation of the fluorescence signal}

In what follows we shall assume that transit time is much larger
than lifetime of the upper state, $\tau _{tr} >> \tau _{nat}$,
which is true for the parameters of our experiment. This allows us
to use the adiabatic elimination for the non-diagonal density
matrix element $\rho _{eg}$ \cite{Stenholm} :

\begin{equation}
\label{eq7} \rho _{eg} \left( {t} \right) = \frac{{i\Omega
\left( {t} \right)}}{{\Gamma _{e} - i2\delta} }\left( {n_{g}
\left( {t} \right) - n_{e} \left( {t} \right)} \right)  \; .
\end{equation}

\noindent The above equation relates $\rho _{eg} \left( {t}
\right) $ to the populations $ n_{g}(t) = \rho _{gg}(t)$ and $n_{e}(t) =
\rho _{ee}(t)$. The decay rate $ \Gamma _{e} = 1/\tau
_{nat}$ gives the natural width of level $e$, while $\delta = 2\pi
\Delta \nu_L$ is the laser detuning. With $\rho _{eg}$ defined by
Eq.~(\ref{eq7}), the time evolution of the populations is given by
simple balance equations:

\begin{eqnarray}
\label{eq8a} \frac{{d}}{{dt}}n_{e} = - \Gamma _{e} n_{e} + r\left(
{t} \right)\left( {n_{g} - n_{e}}  \right); &&\\
\label{eq8b}\frac{{d}}{{dt}}n_{g} = \Pi \Gamma _{e} n_{e} + r\left( {t}
\right)\left( {n_{e} - n_{g}}  \right) . &&
\end{eqnarray}

\noindent The first equation describes the population loss from
level $e$ via two processes: (i) spontaneous decay at the rate $
\Gamma _{e}$, and (ii) stimulated emission at the rate equal to
the optical pumping rate $r\left( {t} \right)$:

\begin{equation}
\label{eq9} r\left( {t} \right) = \Gamma _{e} \frac{{\Omega
^{2}\left( {t} \right)}}{{4\delta ^{2} + \Gamma _{e}^{2}} } \; .
\end{equation}

\noindent The population of level $g$ is affected by three
competing processes:  (i) photon absorption at the rate $r\left(
{t} \right)$) resulting in the population of level $e$, (ii)
return of population from level $e$ to level $g$ due to stimulated
emission, and (iii) return of population from level $e$ to level
$g$ due to spontaneous emission. The rate of the latter is
determined by the branching coefficient $\Pi$ of the given HF
transition. The branching coefficients are normalized such that
$\Pi$=0 for an entirely open system (no spontaneous return from
level $e$ to level $g$) and $ \Pi$=1 for a closed system (no
transitions outside outside the $g-e$ system).

The initial conditions of Eqs.~(\ref{eq8a}) and (\ref{eq8b})
follow from the requirement that initially all the population is
in level $g$ while level $e$ is not populated: $ n_{g} \left( {t =
- \infty}  \right) = 1$; $n_{e} \left( {t = -\infty } \right) =
0$. The assumption $\tau _{tr} > \tau _{nat}$ leads to a further
simplification of Eqs. (\ref{eq8a}) and (\ref{eq8b}) in the weak
excitation limit, when $r\left( {t} \right) < 0.5\Gamma _{e}$. As
weak excitation we understand excitation at laser intensities
smaller than the saturation intensity given by Eq.~(\ref{eq4}),
i.e., when Rabi frequency of the transition does not exceed the
saturated value, $\Omega _{0}<\Omega _{sat}$, where $\Omega _{sat}
\equiv \Gamma_{e}/\sqrt {2}$  \cite{Metcalf}. In that case, the
adiabatic elimination implies that $dn_{e}/dt$=0 \cite{Stenholm},
and Eq. (\ref{eq8a}) immediately yields

\begin{eqnarray}
\label{eq11}
 - \Gamma _{e} n_{e} + r\left( {t} \right)\left( {n_{g} - n_{e}}  \right) =
0
 \Rightarrow \qquad \qquad &&\nonumber\\  \Rightarrow\quad n_{e} \left( {t} \right)
 = \frac{{r\left( {t}
\right)}}{{\Gamma _{e} }}\left( {n_{g} \left( {t} \right) - n_{e}
\left( {t} \right)} \right)&&.
\end{eqnarray}

\noindent
Equation (\ref{eq8b}) can then be transformed into the form

\begin{eqnarray}
\label{eq12} \frac{{d}}{{dt}}\left( {\left( {1 +
\frac{{r}}{{\Gamma _{e}} }} \right)n_{ - }}  \right) = - r\left(
{t} \right)\left( {1 - \Pi}  \right)_{} n_{ -} \left( {t}
\right); &&\nonumber\\  n_{ -}  \left( {t} \right) \equiv n_{g}
\left( {t} \right) - n_{e} \left( {t} \right)&&.
\end{eqnarray}

\noindent The above equation can be comfortably used for the
evaluation of the fluorescence signal $J$. Integration of both
sides of Eq. (\ref{eq8a}) yields the total number of spontaneous
photons emitted by the excited atoms:

\begin{equation}
\label{eq10} J = \Gamma _{e} \int\limits_{ - \infty} ^{\infty}
{dt} n_{e} \left( {t} \right) = \int\limits_{ - \infty} ^{\infty}
{dt} r\left( {t} \right) n_{-}.
\end{equation}

\noindent Integration of Eq.~(\ref{eq12}) and combination of the
result with (\ref{eq10}) yields

\begin{equation}
\label{eq13} \left( {1 - \Pi}  \right)J = 1 - n_{ -}  \left( {t =
\infty}  \right) = 1 - n_{g} \left( {t = \infty}  \right).
\end{equation}

\noindent The above expression has a straightforward physical
meaning: the number of spontaneously emitted photons on
transitions outside the $g-e$ system is equal to the total loss of
ground state population during interaction with the laser field.

Using Eqs.~(\ref{eq12}) and (\ref{eq13}), we can derive the
fluorescence signal in an explicit analytical form:

\begin{eqnarray}
\label{eq14} J = \frac{{1}}{{\left( {1 - \Pi}  \right)}}\left[
{1 - \mathrm{\exp}\left( { - \left( {1 - \Pi}  \right)R}
\right)} \right] \; ;&&\nonumber\\ \quad R = \int\limits_{ -
\infty} ^{\infty} {dt} \frac{{r\left( {t} \right)}}{{1 +
r\left( {t} \right)/\Gamma _{e}}} \; .
\end{eqnarray}

\noindent Since we consider the case of weak excitation, when
$r\left( {t} \right) < 0.5\Gamma _{e} $, the integral \textit{R}
in Eq.~(\ref{eq14}) further simplifies to the form

\begin{eqnarray}
\label{eq15} R \cong \int\limits_{ - \infty} ^{\infty}  {dt}
r\left( {t} \right) = \frac{{\sqrt {\pi}  \Gamma _{e} \tau
_{tr}} }{{2}}\frac{{\Omega _{0}^{2} }}{{4\delta ^{2} +
\Gamma _{e}^{2}} } \; ; &&\nonumber\\ \quad \Omega _{0} < \Omega
_{sat}\equiv \Gamma _{e} /\sqrt{2}.&&
\end{eqnarray}

\noindent Dependence of the fluorescence signal on the laser
detuning $ \delta = 2 \pi \Delta \nu_L$ can now be rewritten as

\begin{eqnarray}
\label{eq16} J\left( {\delta}  \right) = \frac{{\sqrt {\pi}
\Omega _{0}^{2}} \tau _{tr}}{{2\Gamma _{e}}
}\frac{{1}}{{P_{pump}} }\left[ {1 - \mathrm{exp}\left( { -
\frac{{P_{pump}} }{{1 + 4\left( {\delta /\Gamma _{e}}
\right)^{2}}}} \right)} \right] \; ; &&\nonumber\\ \quad P_{pump}
= \frac{{\tau _{tr}} }{{\tau _{pump}^{\left( {0} \right)}} } \; ;
\quad \tau _{pump}^{\left( {0} \right)} = \frac{{2\Gamma _{e}}
}{{\sqrt {\pi} \Omega _{0}^{2} \left( {1 - \Pi}  \right)}} \;
.\quad &&
\end{eqnarray}

\noindent The above equation shows that excitation spectrum
strongly depends on the pumping parameter $P_{pump}$, which is
given by the ratio of transit time $\tau _{tr}$ and pumping time
$\tau _{pump}^{\left( {0} \right)}$. The latter was already
discussed in Sect.~I [Eq.~(\ref{eq2})], and it has the meaning of
optical pumping time at resonant excitation ($\delta $=0).
Importantly, the parameter $P_{pump}$ can be large even at laser
intensities well below the saturation limit: $P_{pump} \gg 1$ when
$\tau _{tr} \gg \tau _{pum}^{\left( {0} \right)}$ and $\Omega _{0}
\ll \Omega _{sat}$.

\subsection{Line broadening by optical pumping}

When the pumping parameter is small ($P_{pump} \ll 1$), equation
(\ref{eq16}) simplifies to yield the ordinary Lorentz lineshapes:

\begin{equation}
\label{eq17} J_{L} \left( {\delta}  \right) = R=\frac{{\sqrt {\pi}
\Omega _{0}^{2}} \tau _{tr}}{{2\Gamma _{e}} }\frac{{1}}{{1 +
4\left( {\delta /\Gamma _{e}}  \right)^{2}}}.
\end{equation}

\noindent When $P_{pump}$ is increased, Eq. (\ref{eq17}) no longer
holds and Eq. (\ref{eq16}) must be used. An almost 10-fold
increase of the linewidth is observed as $P_{pump}$ is increased
form 0.1 to 50 (see Fig.~7). Such broadening has a simple
explanation. Consider a near resonant case, when $\Delta
\nu_L=\delta /2\pi \approx 0$. Starting from values
$P_{pump}\approx 1$ the atoms spend sufficient time in the laser
filed for the population of the ground state to be depleted,
$n_{g} \left( {t = \infty} \right) \simeq 0$. Depletion of level
$g$ is associated with the emission of a fixed number of photons
1/$\left( {1 - \Pi } \right)$ (see Eq.~(\ref{eq13})). Hence,
optical pumping saturates the observed signal $I\left(
{\delta\approx 0} \right)$ via depletion saturation, provided that
$\tau _{pump} \left( {\delta} \right) > \tau _{tr} $. Further
increase of $P_{pump} $ cannot increase the number of photons
emitted upon excitation at the line center. At the same time, the
number of photons emitted upon excitation in the wings continues
increasing with $P_{pump}$ until depletion saturation is reached
at consecutively larger laser detunings $\delta$. Therefore
linewidths in the excitation spectra will increase with
$P_{pump}$, and lineshapes will exhibit the characteristic
flat-top peaks at large $P_{pump}$.

The relative increase of the width $\Delta \nu_{OP} = \delta_{OP}
/ 2\pi$ (FWHM) of the line profile affected by optical pumping as
compared to the natural width can be easily obtained from Eq.
(\ref{eq16}):

\begin{eqnarray}
\label{eq18} \frac{{\Delta \nu _{OP}} }{{\Delta \nu _{nat}} } =
\sqrt {\frac{{P_{pump} }}{{\ln2 - \ln\left( {1 +
\mathrm{\exp}\left( {
- P_{pump}}  \right)} \right)}} - 1} \; ;&&\nonumber\\
 \quad \Delta \nu_{nat} = \Gamma _{e}/ 2\pi \; .&&
\end{eqnarray}

\noindent Variation of the width with $P_{pump}$ is shown on
Fig.~8. As can be seen, the broadening becomes noticeable at about
$P_{pump}=1$, and further increase of the width scales
as square root of $P_{pump}$ for large values of $P_{pump}$.

The condition $P_{pump}>1$ for broadening by optical pumping can
be reformulated in terms of Rabi frequencies, i.e., Rabi frequency
of the laser-driven transition must be larger than some critical
value $\Omega_{cr}$:

\begin{eqnarray}
\label{eq19} \Omega > \Omega _{cr} = \sqrt {\frac{2}
{\pi^{1/2} \cdot \tau _{nat} \tau _{tr}
\left({1 - \Pi}\right)}}
\; ; \\
\Omega _{cr} = \Omega_{sat} \sqrt {\frac{2\tau_{nat}}
{\tau _{tr} \left({1 - \Pi}\right)}} \; .
\end{eqnarray}

\noindent Inequality (\ref{eq19}) generalizes the results obtained
in \cite{Auzinsh, Ruth} for the limit of entirely opened level
systems with $\Pi = 0$, which is often used as an approximation in
the case of molecules with many possible rovibronic transitions.
Broadening by optical pumping turns out to be sensitively
dependent of the branching ratio $\Pi$. In the limit of closed
level system ($\Pi = 1$) $\Omega_{cr}$ is formally equal to
infinity. It is not surprising: if there are two isolated quantum
states, then no pumping can occur regardless how strong is the
exciting laser field.

It is important to note that the above described broadening
mechanism differs from the classical textbook examples of
saturation and power broadening \cite{Demtroder}. In
\cite{Demtroder}, the saturation broadening is attributed to the
strong field effects, when light-induced pumping rate becomes
comparable to the relaxation rates, while the power broadening is
attributed to a considerable Rabi flopping frequency. The
saturation parameter in both cases is given by the ratio of
pumping rate to relaxation rate. In our considered case, in
contrast, the pumping rate is very small, laser intensity is well
below the traditional saturation intensity given by Eq.
(\ref{eq4}), yet a notable depletion of level $g$ is reached via
optical pumping in a partially open two-level system due to long
interaction time with the laser field; the line broadening thus
occurs in the weak excitation limit, when $n_{e} \left( {t}
\right) \ll n_{g} \left( {t} \right)$.

Line broadening by optical pumping is thus the dominating line
broadening mechanism when $\Omega_{crit}<\Omega < \Omega_{sat}$.
At $\Omega> \Omega_{sat}$ the rate of laser-induced transitions
exceeds the spontaneous transition rate. When the splitting ($\sim
\Omega$) of laser dressed states exceeds their widths, the
population is equally shared between the levels $e$ and $g$
\cite{Tannoudji2,Tannoudji3}. The pumping time $\tau _{pump}
\left( {0} \right)$ then stabilizes at $2/\left(\Gamma _{e}\left(
{1 - \Pi} \right)\right)$ and becomes independent on further
increase of laser intensity.


\begin{figure}
\label{Fig7}
\includegraphics[width=7cm]{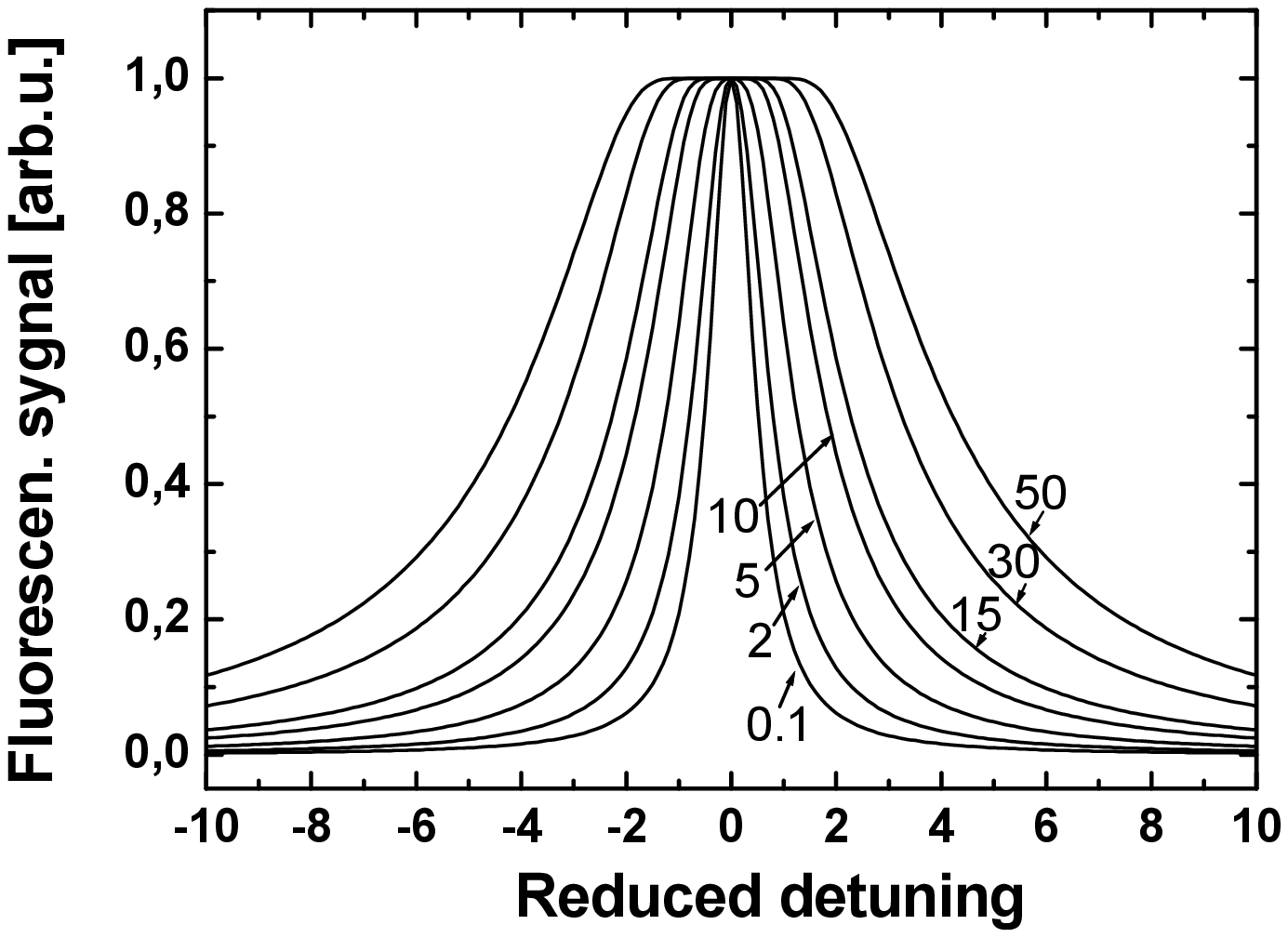}
\caption {Signal $J$ (Eq.~(\ref{eq16})) as a function of the
reduced detuning $\delta /\Gamma _{e}$ for different values of the
pumping parameter $P_{pump}$ (shown as labels of the curves).}
\end{figure}



\begin{figure}
\label{Fig8}
\includegraphics[width=7cm]{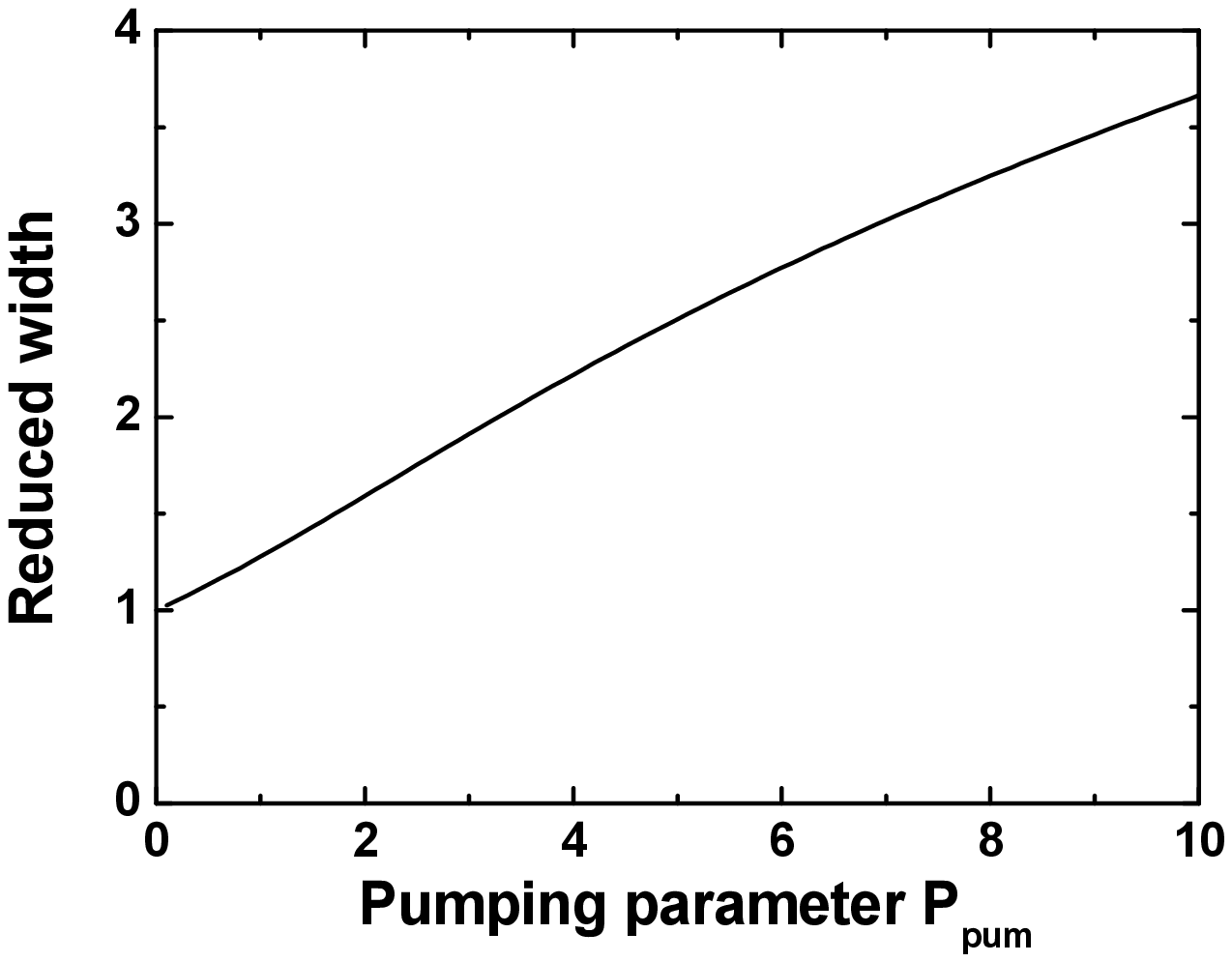}
\caption{The reduced width $\Delta \nu_{pum}/\Delta
\nu_{nat}$ given by Eq.~(\ref{eq18}) as a function of the pumping parameter $P_{pump}$.}
\end{figure}



\subsection{Critical laser intensity}

It is useful to rewrite Eq. (\ref{eq19}) in terms of laser
intensities, since those are usually used by experimentalists. We
shall therefore express the laser intensity in terms of the
transition Rabi frequency as follows:

\begin{equation}
\label{eq20} I_{las} = \frac{{4\pi ^{2}\hbar c}}{{3\lambda
^{3}}}\frac{{\Omega ^{2}}}{{\Gamma _{e} \Pi} }.
\end{equation}

\noindent The above equation is well known for closed systems with
$\Pi$=1 \cite{Metcalf}. In the case of a partially open systems we
have replaced the natural width $\Gamma _{e}$ by the partial
natural width $\Gamma _{e} \Pi$ of the given transition. This can
be done because Rabi frequency $\Omega = Ed/\hbar$ involves the
dipole element $d =\langle g|d_{z} |e\rangle $ associated with the
partial natural broadening $\Gamma_{e} \Pi = 4\omega^{3}
d^{2}/3\hbar c^{3}$ \cite{Sobelman}. Since $I_{las} = E^{2}c/8\pi
$, we obtain Eq. (\ref{eq20}). Using Eq. (\ref{eq20}), equation
(\ref{eq19}) can be rewritten as

\begin{eqnarray}
\label{eq21} I_{cr} = \frac{{8\pi ^{2}\hbar c}}{{\sqrt {\pi}
3\lambda ^{3}}}\frac{{1}}{{\tau _{tr} \Pi \left( {1 - \Pi}
\right)}} = \frac{{4\tau _{nat}} }{{\sqrt {\pi}  \tau _{tr} \left(
{1 - \Pi}  \right)}}I_{sat} \; .
\end{eqnarray}

\noindent Simultaneously, the pumping parameter can now
be rewritten in terms of ratio of actual laser intensity to critical laser intensity:

\begin{equation}
\label{eq23} P_{pum} = I_{las}/I_{cr}.
\end{equation}

\noindent Equation (\ref{eq21}) shows the relation between the
critical laser intensity and the traditional saturation intensity
[Eq. (\ref{eq4})], which gives the limit for onset of power
broadening.

Note, that formally $I_{cr}\rightarrow \infty$ when $\Pi
\rightarrow 0$ (completely open system). The limit of $\Pi
\rightarrow 0$ corresponds to $d\rightarrow 0$, i.e., to forbidden
optical transitions. Therefore, even a very small transfer of
population from level $g$ to level $e$ will require extremely
large laser intensity.




\section{Residual Doppler broadening}

Besides the broadening due to optical pumping, the spectral lines
are also affected by a small but non-negligible Doppler broadening
due to finite divergence angle $\vartheta$ of the atomic beam.
Such divergence is associated with non-zero velocity components in
the direction of the laser beam for atoms moving not exactly
parallel to the atomic beam axis. For atoms experiencing the
Doppler shift $\Delta \omega$ the laser detuning $\delta$ will
transform into the detuning $\delta + \Delta \omega$. A
corresponding replacement $\delta \to \delta + \Delta \omega$
should therefore be done in Eq. (\ref{eq16}). The resultant
profile of a line in the excitation spectrum is thus given by a
sum of profiles (\ref{eq16}) resulting from absorption of laser
photons by atoms of different velocity groups. If the probability
of atoms to have a velocity leading to the Doppler shift $\Delta
\omega$ is $P_{D} \left( {\Delta \omega} \right)$, then the
resultant line profile is given by the integral

\begin{equation}
\label{eq22} J_{res} \left( {\delta}  \right) = \int_{ - \infty}
^{\infty}  {d\Delta \omega}  P_{D} \left( {\Delta \omega}
\right) J \left( {\delta + \Delta \omega}  \right).
\end{equation}

\noindent The analytical form of the function $P_{D} \left(
{\Delta \omega} \right)$ for effusive beams has been derived in
\cite{OptSpectr}, while for the case of supersonic beams it will
be analyzed in detail in \cite{JPB}. This analysis builds on the
following assumptions: (i) nozzle diameter $d$ is small compared
to the diameter $b$ of the entrance aperture and distance $L$ from
the nozzle to the excitation zone (see Fig.~1); (ii) divergence
angle $\vartheta$ of the atomic beam is small; (iii) size of the
excitation zone $\sim r_{las} $ is small compared to the distance
$L$, and the distribution of atoms within $\sim r_{las} $ is
uniform; (iv) the velocity distribution in the direction
perpendicular to the atomic beam axis is due to the divergence of
the beam with the axial velocity distribution $F(v)$. The
distribution functions $F(v)$ for various kinds of beams can be
found in \cite{Ramsey}.

Leaving out the somewhat lengthy detailed derivation of the
function $P_{D} \left( {\Delta \omega} \right)$ to the forthcoming
paper \cite{JPB}, we shall give here the final form of the most
essential core ($|\Delta \nu| < \Delta \nu_D / 1.5$) part of the
distribution function for the supersonic beam:

\begin{equation}
\label{eq32} P_{D}^{\left( {cor} \right)} \left( {\Delta \nu}
\right) = \frac{{2}}{{\pi \Delta \nu _{D}} }\sqrt {1 - \Delta
\nu ^{2}/\Delta \nu _{D}^{2}} \; ,
\end{equation}

\noindent with

\begin{equation}
\label{eq27}  \Delta \nu_D\equiv
\frac{v_f}{\lambda}\frac{\vartheta}{2} \; ; \qquad
\vartheta=\frac{b+d}{L}. \;
\end{equation}
\noindent The values of the parameters $v_f$, $\lambda$, $b$, $d$,
and $L$ are given in Sect. II. Note, that the function
(\ref{eq32}) deviates strongly from the Gaussian function, which
is usually associated with Doppler profiles. The frequency
dependence of $P_D$ in the wings of the spectral line ($|\Delta
\nu| \geq \Delta \nu_D / 1.5$) differs from that given by
Eq.~(\ref{eq32}). Nevertheless, in our case it is sufficient to
use only the core part of $P_D$. Since $\Delta \nu_D$ is
comparable with $\Delta \nu_{nat}$, the natural broadening
outcompetes the exponentially small wings of the Doppler profile
at large $\Delta \nu$ \cite{JPB}.




\section{Results and discussion}

Calculations of the theoretical spectra are performed in two
steps: (i) solution of the evolution problem for an individual
atom excited by linearly polarized laser field detuned by $\Delta
\nu = \delta / 2\pi$, and (ii) calculation of the resultant line
profile by performing the convolution (\ref{eq22}). The first step
is performed by modeling quantum dynamics of individual pairs of
Zeeman sublevels $m_F$ within the $F''m_{F''}\rightarrow F'm_{F'}$ HF
transition during coupling of the levels by the electrical field
of laser light distributed as  $\mid \vec{E}\mid =E_0
\mathrm{\exp}\left( { - z^{2}/2r_{las}^{2}}  \right) \mid
\vec{e}_z \cos (\omega t)\mid $. Correspondingly, the spatial
distribution of Rabi frequencies of individual HF transitions also
follow the Gaussian distribution:

\begin{equation}
\label{eq33a} \Omega^{(m)} = E_0  \exp \left( - z^{2}/2r_{las}^{2}
\right)\langle F''m_{F''}|d_z|F'm_{F'} \rangle \; .
\end{equation}

\noindent It is convenient to introduce the reduced Rabi frequency
$\Omega_{red}$ associated with unresolved $3s-3p$ transition:

\begin{equation}
\label{eq33} \Omega_{red} \equiv \frac{E_0}{\hbar} \sqrt{|( 3s||D||3p )|} \; .
\end{equation}

\noindent Rabi frequencies of individual Zeeman components can
then be calculated from $\Omega_{red}$ using the known line
strengths $\tilde{S}_i^{(j)}$ given in Fig.~6 and the 6j symbols
\cite{Sobelman}:

\begin{equation}
\label{eq34}
\Omega_0^{(m)}=\Omega_{red}\sqrt{\tilde{S}_i^{(j)}}\left(
\begin{array}{lcr}
 F''&1& F' \\
 -m_{F''} &0 &m_{F'}  \\
 \end{array}
\right) \; ,
\end{equation}

\noindent where indexes $i$ and $j$ stay for the chosen HF
component and the chosen $j$ level of the upper state,
respectively.

The values of $\Omega_{red}$ used in the calculations may be
obtained from their relation to the laser intensity $I_{las}$ [see
Eq. (\ref{eq20})]:

\begin{equation}
\label{eq36} I_{las}=\frac{4 \pi^2 \hbar c }{3 \lambda^3}
\frac{\tau_{3p}\Omega^2_{red}}{g_{3p}} \; ,
\end{equation}

\noindent where $g_{3p}=3$ is the statistical weight of the $3p$
state. Note, that in the calculations of theoretical spectra we
used $\Omega_{red}$ as the only fitting parameter. The theoretical
values of $I_{las}$ given in Figs.~3-5 were calculated from the
fitted $\Omega_{red}$ values using Eq.~(\ref{eq36}), and they are
in a good agreement with the experimental values calculated from
measured laser power and radius of the laser beam.

For a qualitative interpretation of the experimental results it is
helpful to consider a simplified model, in which the populations
of Zeeman components evolve independently, and the resulting
signal  $J$ is simply a sum of individual signals $J_m$ with the
same $m_{F''}=m_{F'}=m$: $J(\Delta \nu)=\sum\limits_m J_m(\Delta
\nu)$. In the reality, however, Zeeman sublevels are subject to
the spontaneous emission on transitions with $\Delta m_F = \pm 1$.
As will be shown below, such cascading can significantly change
the excitation spectrum as compared to the simplified treatment
when couplings with $\Delta m_F = \pm 1$ are neglected.

In order to
account for cascading, we have elaborated an accurate numerical
algorithm allowing the integration of equations of motion for
the density matrix \cite{Tannoudji1, Shore}

\begin{equation}
\label{eq35}
\frac{d\rho}{dt}=-\frac{i}{\hbar}[H,\rho]-\frac{1}{2}(\Gamma\rho+\rho\Gamma)+L(\rho),
\end{equation}

\noindent whereby Zeeman structure of all sublevels of the system
depicted in Fig.~2 is taken into account. In equation
(\ref{eq35}), the Hamiltonian $H$ describes the system "atom +
laser field", the matrix $\Gamma$ describes the spontaneous
emission, and $L(\rho)$ describes the cascade effects and has a
simple explicit form in the representation of polarization moments
\cite{Auzinsh}. In order to achieve a fast and efficient solution
of Eq.~(\ref{eq35}), we employ the split propagation technique
\cite{fiet, andrei}.

\subsection{Regular changes of line profiles}

\begin{figure}
\label{Fig9}
\includegraphics[width=7cm]{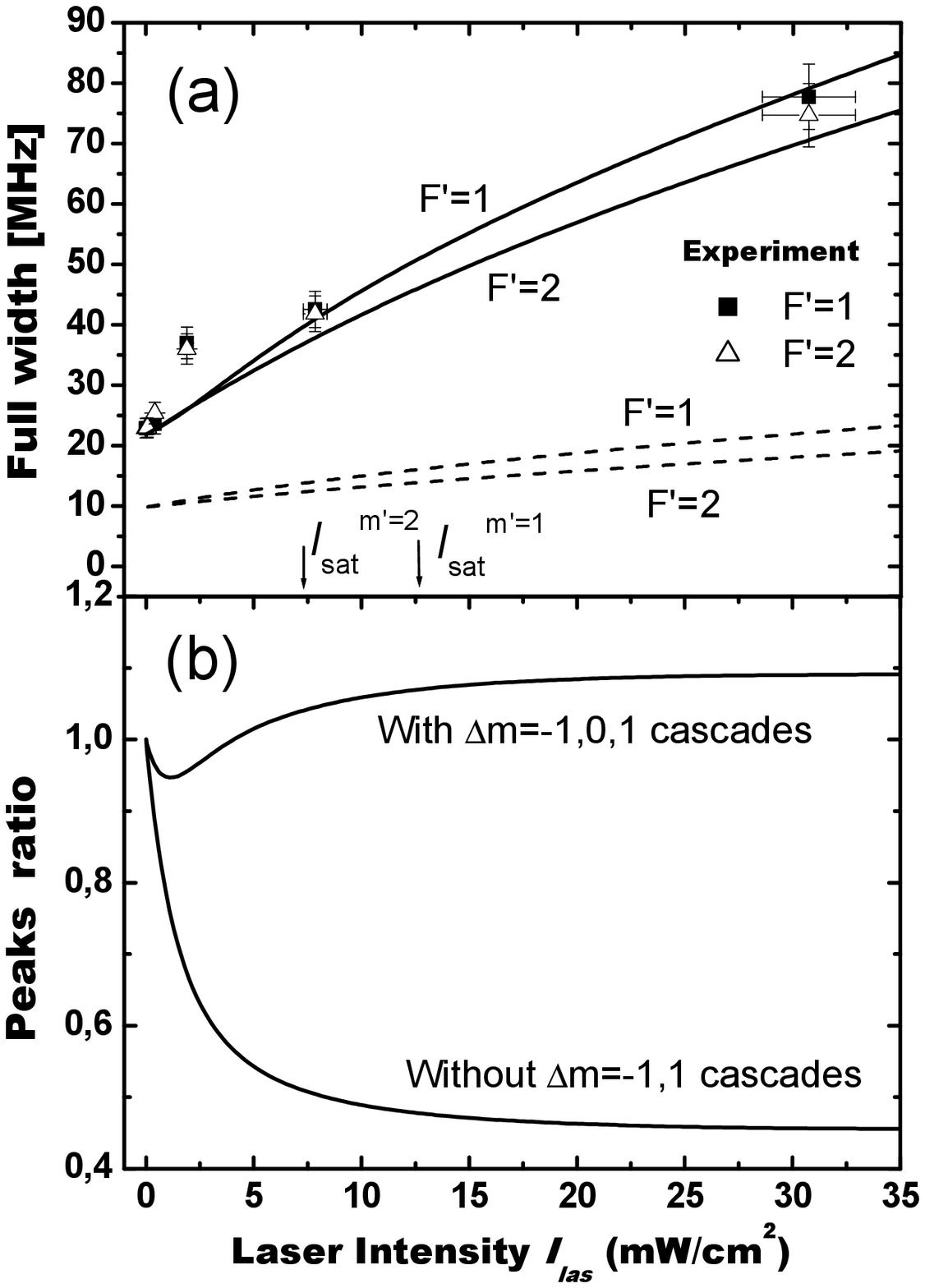}
\caption{(a) The FWHM widths of the lhs ($F'=1$) and rhs ($F'=2$)
peaks of Fig.~3 as a function of laser intensity $I_{las}$. Solid
squares - experiment, $F'=1$; open triangles - experiment, $F'=2$;
solid curves - theory; dashed curves - pure power broadening
neglecting the broadening due to optical pumping. Arrows indicate
saturation intensities of both transitions. (b) Calculated peak
ratio $\Re$ of the rhs ($F'=2$) and the lhs ($F'=1$) peaks of
Fig.~3 as a function of laser intensity $I_{las}$. The calculation
was performed with and without taking the $\Delta m_F = \pm 1$
cascades into account.}
\end{figure}

\begin{figure*}
\label{Fig10}
\includegraphics[width=14cm]{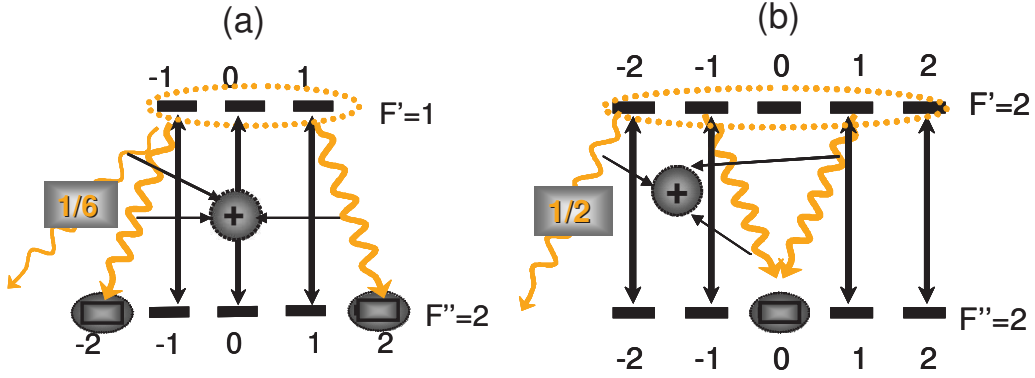}
\caption{Zeeman sublevels involved in (a) the $3s_{1/2},F''=2
\rightarrow 3p_{1/2}, F'=1$ transition, and (b) the
$3s_{1/2},F''=2 \rightarrow 3p_{1/2}, F'=2$ transition.
Spontaneous emission leads to the population loss to the $F''=1$
level of the ground state, and to the dark $m_{F''}=\pm 2$ levels
in (a) and to $m_{F''}= 0$ in (b). The effective branching
coefficient in case (a) is therefore $\Pi_{eff}(2,1)<5/6$ and in
case (b) it is $\Pi_{eff}(2,2)<1/2$}.
\end{figure*}


The calculated excitation spectra $J(\Delta \nu)$ in the case of
very small laser intensities ($I_{las} < I_{cr}$) are shown in
Figs.~3(a), 4(a), and 5, whereby the residual Doppler broadening
has been taken into account by performing the convolution
(\ref{eq22}). An excellent agreement with the experimental results
is observed. One can also see that the relative peak intensities
correspond to those expected from the theoretical line strengths
given in Fig.~6.

The theoretical spectra in the case when $I_{cr} < I_{las} <
I_{sat}$ are shown in Figs.~3(b), 4(b), and 5. An interesting
observation can be made in Fig. 3(b). Intuitively, one would
expect that optical pumping is manifested more strongly and at
smaller laser intensities for lines with smaller values of
branching ratio $\Pi_i$, when only a small fraction of population
spontaneously returns to the initial level. This is, however, not
the case. In fact, Eq.~(\ref{eq21}) implies that the critical
laser intensity $I_{cr}$  has a minimum at $\Pi=0.5$. Therefore,
nonlinear effects associated with optical pumping are more
pronounced for HF transitions with branching coefficients $\Pi$
close to 0.5. Note, that all the excited HF levels considered here
have equal lifetimes (16.2~ns). Low values of $\Pi_i^{(j)}$ are
thus associated with low values of both the line strengths
$S_i^{(j)}$ and the individual Rabi frequencies
$\Omega_{0}^{(m)}$, such that interaction with laser light is
inefficient for transitions with small $\Pi_i^{(j)}$. Optical
pumping turns out to be most pronounced for transitions with
$\Pi=0.5$. This can be best seen in Fig.~5: the relative intensity
of the smaller peak with $\Pi_i = 1/6$ increases with respect to
the stronger peak with $\Pi_i = 1/2$ as the laser intensity is
increased  in the range $I_{cr} < I_{las} < I_{sat}$. This is
because the lower level in the case of component with $\Pi_i =
1/2$ is faster depleted than it is in the case of component with
$\Pi_i = 1/6$.

Another important consequence of optical pumping is line
broadening, which can be observed when laser intensity $I_{las}$
is close to the critical value $I_{cr}$ [Eq.~(23)] of the given
transition, or when Rabi frequency $\Omega$ is close to the
critical Rabi frequency $\Omega_{cr}$ given by Eq.(21). Since
$I_{cr}$ has a minimum at $\Pi=0.5$, the spectral lines with such
branching ratio are most strongly affected by broadening due to
optical pumping. Dependence of the linewidth on laser intensity is
illustrated in Fig.~9(a) for the $3s_{1/2}, F''=2 \rightarrow
3p_{1/2}, F'=1$ transition with $\Pi_i=5/6$ (lhs peak in Fig.~3)
and the $3s_{1/2}, F''=2 \rightarrow 3p_{1/2}, F'=2$ transition
with $\Pi_i=1/2$ (rhs peak in Fig.~3). One can see that remarkable
broadening takes place at laser intensities below the saturation
intensity $I_{sat}$ [marked in Fig.~9(a) with arrows], while the
component with $\Pi_i=1/2$ ($F'=2$) exhibits broadening at smaller
intensities than the other transition. For comparison, dashed
curves in Fig.~9(a) show the intensity dependence of the power
broadened linewidths calculated as $\Delta \nu _{pow}\! =\! \Delta
\nu _{nat}\sqrt {1 + I^{2}/I_{sat}^{(i)2} }$ \cite{Metcalf}. It is
immediately obvious that in the laser intensity range considered
here the power broadening is much smaller than broadening due to
optical pumping even at intensities exceeding the saturation
intensity.

\subsection{Irregular changes of line profiles and Zeeman structure}

Variations of the excitation spectrum of the $ 3s_{1/2}, F''=2
\rightarrow 3p_{1/2}, F'=1,2$ transitions with laser intensity
(Fig.~3) are significantly different from those observed for the
$3s_{1/2}, F''=1 \rightarrow 3p_{1/2}, F'=1,2$ transition (Fig.~5)
in two ways: (i) relative intensity of the peak with $\Pi_i=1/2$
first decreases slightly and then increases as the laser intensity
is increased; (ii) Broadening of the peak with $\Pi_i=1/2$ is
actually smaller than broadening of the peak with $\Pi_i=5/6$. The
key of understanding such striking differences is in the different
Zeeman sublevel structure in both cases. Numerical simulations
using Eq.~(\ref{eq35}), which include cascade transitions with
$\Delta m_F = \pm 1$, yield a ratio $\Re$ between the peak with
$\Pi_i=1/2$ and the peak with $\Pi_i=5/6$, which initially
decreases with increasing laser intensity and reaches a minimum at
$I_{las} \approx  1.3~\textrm{mW}/\textrm{cm}^2$ [see Fig.~9(b)].
As laser intensity is further increased, the ratio starts growing,
reaches unity at $I_{las} \approx  3.2~\textrm{mW}/\textrm{cm}^2$,
and grows to values slightly larger one. If the cascade
transitions with $\Delta m_F = \pm 1$ are ignored, the
calculations yield a monotonously decreasing ratio $\Re$ [lower
curve in Fig.~9(a)] without any "abnormalities".

The effect of $\Delta m_F = \pm 1$ transitions becomes obvious at
closer inspection of Zeeman sublevels involved in the $3s_{1/2},
F''=2 \rightarrow 3p_{1/2}, F'=1$ [Fig.~10(a)] and $3s_{1/2},
F''=2 \rightarrow 3p_{1/2}, F'=2$ [Fig.~10(b)] transitions. Since
the laser field is linearly polarized, only the levels with the
same $m_F$ are coupled by it. The presence of "dark" levels
becomes immediately obvious. In the case of the $F''=2 \rightarrow
F'=1$ transition, the $m_F=\pm 2$ sublevels of the lower level are
not coupled by the laser field and thus act as dark states, which
accumulate population channeled to them vie optical pumping from
the $m_F=\pm 1$ sublevels of the upper level. As a result, the
branching coefficient $\Pi_i=5/6$ should be replaced by a smaller
effective branching coefficient $\Pi_{eff}(F''=2,F'=1) < 5/6$,
which accounts for the population loss to dark states. In the case
of the $F''=2 \rightarrow F'=2$ transition the dark state is
$m_F=0$ (due to the selection rule $\Delta m_F \neq 0$ for
$F''=F'$), therefore $\Pi_{eff}(2,2) < 1/2$.

Increase of laser intensity leads to a larger population of the
dark states, and, consequently, to a monotonous decrease of
$\Pi_{eff}(F'',F')$. At very weak laser fields
$\Pi_{eff}(2,1)=5/6$,  $\Pi_{eff}(2,2)=1/2$, and
$I_{cr}(2,1)>I_{cr}(2,2)$ [see Eq.~(\ref{eq21})]. Hence, the
transition $F''=2 \rightarrow F'=2$ is more strongly affected by
optical pumping than the other transition. Correspondingly, the
ratio $\Re$ decreases with increasing laser intensity. As laser
intensity is further increased, both $\Pi_{eff}(2,1)$ and
$\Pi_{eff}(2,2)$ decrease. At some value of $I_{las}$ both
effective branching ratios satisfy the equality
$\Pi_{eff}(2,1)(1-\Pi_{eff}(2,1))=\Pi_{eff}(2,2)(1-\Pi_{eff}(2,2))$.
In that case, $I_{cr}(2,1)=I_{cr}(2,2)$, and both HF transitions
are equally strongly affected by optical pumping. This corresponds
to the ratio $\Re = 1$ at $I_{las}=I_{eq}=3.2$mW/cm$^2$ in
Fig.~9(b).  At $I_{las}>I_{eq}$ the value of $\Pi_{eff}(2,2)$
becomes larger than $\Pi_{eff}(2,1)$, such that
$I_{cr}(2,1)<I_{cr}(2,2)$, and the ratio $\Re$ becomes larger than
one. At large laser intensities the ratio $\Re$ asymptotically
approaches the value of 1.09. This is because the populations of
the dark $m_{F''}$ levels reach their maximum possible values when
other $m_{F''}$ levels are fully depleted. In the large intensity
limit the values $\Pi_{eff}(2,2)$ and $\Pi_{eff}(2,1)$ differ by
only 9\%, therefore both HF transitions exhibit similar broadening
due to optical pumping (see Fig.~3). In contrast, in the case of
the $3s_{1/2}, F''=1 \rightarrow 3p_{1/2},F'=1,2$ transitions the
component with $\Pi_i=1/2$ is apparently more strongly broadened
than the component with $\Pi_i=1/6$, which could be expected (see
Fig.~5).

\section{Summary}

We have analyzed the effects of line broadening and redistribution
of relative peak intensities in the hyperfine excitation
spectra of Na atoms due to optical pumping in the weak excitation limit, when
interaction times of atoms with the laser field are long compared
to the characteristic optical pumping time. The study was
motivated by the lack of availability of detailed theoretical
models describing such kind of effects in partially open level
systems at laser intensities below the saturation limit. A number
of significant results were obtained: (i) it is shown that
spectral lines can be significantly broadened at laser intensities
well below the saturation intensity, which is usually regarded as
a threshold for onset of broadening effects; (ii) it is shown that
the presence of dark $m_F$ sublevels can vary the effective
branching coefficients of the transitions, and this variation
depends on laser intensity. Changes in the effective branching
coefficients lead to irregular changes of peak ratios, like
minimum in the intensity dependence of the peak ratio, which deviate
from those expected from the given original branching
coefficients; (iii) analytical expressions are derived, which
allow the calculation of critical values for laser intensity and
Rabi frequency, above which linewidths and peak ratios are notably
affected by optical pumping; (iv) it is shown that the critical
laser intensity and critical Rabi frequency depend on the
branching coefficient $\Pi$ of the transition, and they have a
minimum at $\Pi = 1/2$.

Accurate theoretical simulations of the density matrix equations
of motion using the split propagation technique yielded a good
agreement with the experimental observations. In this study we
have explored the limiting case of long interaction times of atoms
with laser field, which justified the use of the adiabatic
elimination approach. It is possible, however, to obtain explicit
formulas for the excitation spectra in the weak excitation limit
also without the limitation of adiabaticity in switching Gaussian
laser pulses. In the forthcoming publication we shall discuss some
unexpected effects related to transit time broadening in the other
limiting case, when the transit time is much smaller than the
natural lifetime.




\begin{acknowledgments}

This work was supported by the EU FP6 TOK Project LAMOL (Contract
MTKD-CT-2004-014228), NATO Grant EAP.RIG.981378, RFBR Grant No.
08-02-00136, INTAS Young Scientist Fellowship 04-83-3692, Latvian
Science Council, and European Social Fund. We thank profs. K.
Bergmaan, H. Metcalf, and M. Ausinish for helpful discussions.

\end{acknowledgments}



\begin{thebibliography}{10}

\bibitem{Happer} W.~Happer, Rev.~Mod.~Phys. \textbf{44}, 169, (1972).

\bibitem{Tannoudji2} C.~Cohen-Tannoudji, Rev.~Mod.~Phys. \textbf{70}, 707, (1998).


\bibitem{Metcalf} H.~J.~Metcalf, P.~van~der~Straten, \textit{Laser
Cooling and Trapping} (Springer-Verlag, New York, 1999).


\bibitem{Bergmann} K.~Bergmann, U.~Hefter, and J.~Witt, J.~Chem.~Phys.
\textbf{72}, 4777 (1980); H.~M.~Keller, M.~K\"{u}lz, R.~Setzkorn,
G.~Z.~He, K.~Bergmann, and H.~G.~Rubahn,
J.~Chem.~Phys.\textbf{96}, 8819 (1992).


\bibitem{Auzinsh} M.~Auzinsh and R.~Ferber, \textit{Optical Polarization
of Molecules }(Cambridge University Press, Cambridge, U.K., 1995).

\bibitem{Demtroder} W.~Demtr\"{o}der, \textit{Laser Spectroscopy}
(Springer, Berlin, 2003).


\bibitem{Ivanov} V.~V.~Ivanov, \textit{Transfer of radiation in spectral
lines}, NBS Special Publication No. 385 (U.S. GPO, Washington,
1973).

\bibitem{BezAig} N.~N.~Bezuglov, A.~Ekers, O.~Kaufmann, K.~Bergmann,
F.~Fuso, and M.~Allegrini, J.~Chem.~Phys., \textbf{119}, 7094,
(2003).

\bibitem{Dep1} R.~M.~Jopson, R.~R.~Freeman, W.~E.~Cooke, J.~Bokor,
Phys. Rev. A. \textbf{29}, 3154 (1984); A.~Nussenzweig,
E.~E.~Eyler, T.~Bergeman, E.~Pollack, Phys.~Rev.~A. \textbf{41},
4944 (1990).

\bibitem{Dep2} R.~C.~Ekey and E.~F.~McCormack, J.~Phys.~B:
 \textbf{38}, 1029 (2005).

\bibitem{Tannoudji1} C.~Cohen-Tannoudji, G.~Grynberg, and J.~Dupont-Roc,
\textit{Atom-Photon Interactions: Basic Processes and
Applications} (Wiley, New York, 1998).

\bibitem{Ruth} R.~Garcia-Fernandez, A.~Ekers, J.~Klavins,
L.P.~Yatsenko, N.N.~Bezuglov, B.W.~Shore, and K.~Bergmann,
Phys.Rev.A. \textbf{71}, 023401 (2005).

\bibitem{Jones}  K.~M.~Jones, P.~S.~Julienne, P.~D.~Lett, W.~D.~Phillips,
E.~Tiesinga, and C.~J.~Williams, Europhys. Lett. \textbf{35}, 85
(1996).

\bibitem{Sobelman} I.~I.~Sobel'man, \textit{Atomic Spectra and
Radiative Transitions }(Nauka, Moscow, 1977; Springer, Berlin,
1999).

\bibitem{Stenholm} S.~Stenholm, \textit{Foundations of Laser
Spectroscopy} (Wiley, New York, 1984).

\bibitem{Tannoudji3} J.~Dalibard and C.~Cohen-Tannoudji, J. Opt. Soc.
Am. B \textbf{2}, 1707 (1985).

\bibitem{OptSpectr} N.~N.~Bezuglov, M.~Zakharov, A. N.~Klyacharev A.~Ekers,
A. A.~Matveev, K.~Miculis, E.~Saks, I.~Sydoryk, and A.~Ekers,
Opt. Spectrosc. \textbf{102}, 819, (2007) (Opt. Spectrosk. \textbf{102}, 893, (2007)).

\bibitem{JPB} N.~N.~Bezuglov, I.~I.~Beterov, A.~Ekers, K.~Miculis,
E.~Saks, A.~Janovs, P.~Spels, I.~Sydoryk, M. Yu. Zaharov (unpublished).

\bibitem{Ramsey}  N.~F.~Ramsey, \emph{Molecular Beams} (Clarendon,
Oxford, 1989).

\bibitem{Shore} B.W.~Shore,
\textit{The Theory of Coherent Atomic Excitation}(Wiley, New York,
1990).

\bibitem{fiet}
M.~D.~Fiet, J.~A. Fleck, and A.~Steiger, J.~Comput.~Phys.
\textbf{47}, 412 (1982).

\bibitem{andrei}
A.~K.~Kazansky, N.~N.~Bezuglov, A.~F.~Molisch, F.~Fuso, and
M.~Allegrini, Phys.~Rev.~A \textbf{64}, 022719 (2001).


\end{thebibliography}
\end{document}